\documentclass{eas}
\usepackage{graphicx}
\newcommand\langeditorchanges[1]{#1}

\newcommand\edchange[1]{#1}
\newcommand\ednote[1]{}

\newcommand\mathrmpl{_{\langeditorchanges{{\mathrm p}}}}

\usepackage[authoryear]{natbib}

\newcommand{\apj}{{ApJ}}
\newcommand{\apjl}{{ApJL}}
\newcommand{\aap}{{A\&A}}
\newcommand{\aaps}{{A\&A Suppl.}}
\newcommand{\mnras}{{MNRAS}}

\newcommand{\gsim}{\raisebox{-0.6ex}{$\stackrel{{\displaystyle>}}{\sim}$}}
\TitreGlobal{Extrasolar Planets in Multi-body Systems: Theory and Observations}
\Editors{Editors: K. Go\'zdziewski, A. Niedzielski and J. Schneider}
\begin{document}

\title{The formation of massive planets in binary \langeditorchanges{star systems}} 
\author{Wilhelm Kley}\address{Institute of Astronomy \& Astrophysics, University of T\"ubingen, Germany}
%
%
\runningtitle{\edchange{W. Kley: Formation of planets in binary star systems}} 
\begin{abstract}

As of today over 40 planetary systems have been discovered in
binary  \langeditorchanges{star systems}. 
%
%
In all cases the configuration appears to be circumstellar,
where the planets orbit around one of the stars, the secondary acting as
a perturber.  The formation of planets in binary \langeditorchanges{star systems} is more difficult
than around single stars due to the gravitational action of the companion
on the dynamics of the protoplanetary disk.
In this contribution we first  \langeditorchanges{briefly} present the relevant observational
evidence  \langeditorchanges{for} planets in binary systems. Then the dynamical influence that a
secondary companion has on a circumstellar disk will  \langeditorchanges{be} analyzed through
\langeditorchanges{fully} hydrodynamical simulations. We demonstrate that the disk becomes
eccentric and shows a coherent precession around the primary star.
Finally, fully hydrodynamical simulations of evolving protoplanets
embedded in disks  in binary \langeditorchanges{star systems} are presented.
We investigate how the orbital evolution of protoplanetary embryos
and their mass growth from cores to
massive planets might be affected in this very dynamical environment.
We  \langeditorchanges{consider, in particular,} the planet
orbiting the primary in the system $\gamma$ Cephei.
\end{abstract}

\maketitle

%
\section{Observational Data}

Planet formation is obviously a process that occurs around single as well as
in multiple \langeditorchanges{star systems}, a fact that is indicated by the detection of well over 40 planetary
systems that reside in a binary or even multiple star configurations.
All of the observed systems display a so called S-type configuration in which the planets orbit
around one of the stars and the additional star, the companion or secondary star,
acts as a perturber to this system. In this review we shall refer to the secondaries as
single objects, even though they may be multiple.
As indicated in Table~\ref{kley-tab:systems} the distances of the secondaries \langeditorchanges{from} the host
\langeditorchanges{stars} of the planetary \langeditorchanges{systems} range  from very
small values of about 20 AU for Gl~86 and $\gamma$ Cep to several thousand AU.
There are now 4 confirmed systems with a binary separation in the
order of 20 AU; in \langeditorchanges{Table~}\ref{kley-tab:systems} \langeditorchanges{these 
are} shown below the horizontal separation line.
The mere existence of these 4 \langeditorchanges{systems represents} a special challenge to
any kind of planet formation process, due their tightness.
Interestingly, there appears to be a lack of planets for \langeditorchanges{intermediate separations} as there
are no planets in binaries with separations between 20 and 100 AU.
There are many more systems with larger \langeditorchanges{separations} (not listed in the table),
but in most of the cases only projected distances
can be given, and the real physical \langeditorchanges{separations are necessarily} larger. 

\begin{table}[h,t]
\label{kley-tab:systems}
\begin{center}
\begin{tabular}{|l|l|l|l|l|l|} \hline
Star  & $a_\mathrm{bin}$ {\small{[AU]}}
      &  $a_\mathrm{p}$  {\small{[AU]}}
  & $M\mathrmpl\sin$ i  {\small{ [M$_{\rm Jup}$]}}
   &  $e_\mathrm{p}$ 
   & Remarks \\
 \hline
HD 40979  & 6400   &  0.811 & ~3.32  & .23 & \\
Gl 777 A  & 3000   &  3.65  & ~1.15  & .48 & \\
HD 80606  & 1200   &  0.439 & ~3.41  & .93 & \\
55 Cnc B  & 1065   &  0.1-5.9 & ~0.8-4.05  & .02-.34 & {\small multiple} \\
16 Cyg B  & 850    &  1.66  &  ~1.64  & .63  & \\
$\upsilon$ And  & 750   &  0.06-2.5 &  0.7-4.0  & .01-.27 &  {\small multiple} \\
HD 178911 B  & 640    &  0.32  & ~6.3   & .12  & \\
HD 219542 B  & 288    &  0.46  & ~0.30  & .32  & \\
$\tau$ Boo  & 240    &  0.05  &  ~4.08  & .02  & \\
HD 195019  & 150    &  0.14  &  ~3.51  & .03  & \\
HD 114762  & 130    &  0.35  &  11.03   & .34 &  \\
HD 19994  & 100    &  1.54  &  ~1.78  & .33  & \\ 
\hline
HD 41004A  & 23     &  1.33  &  ~2.5   & .39 & {\small multiple} \\
$\gamma$ Cep  & 20.2   &  2.04  &  1.60   & .11 & {\small $e_\mathrm{bin}= 0.4$} \\
{HD 196885}  & 17     &  2.63  &  2.96   & .46 & {\small $e_\mathrm{bin}= 0.4$} \\
Gl 86  & 20     &  0.11  &  4.0    & ~0.046 & {\small White Dwarf} \\
 \hline
\end{tabular}
\end{center}
\caption{Some observed planets in binary \langeditorchanges{star systems}. This is a selection with
emphasis on the shorter period binaries (see \,
\cite{2004A&A...417..353E,2006ApJ...646..523R,2008A&A...479..271C}).
\langeditorchanges{The} list is very incomplete for larger separations.
}
\end{table}
Despite the actual detection of planets in binary systems there is additional circumstantial
evidence of debris disks (which are thought to be a byproduct of the planet formation process)
in binary systems as indicated by Spitzer data. Here, for S-type configurations it is
found that disks around an individual star of the binary exist mainly for binary
separations larger than 50 AU, while P-type circumbinary debris disks are detected only in very tight binaries
with $a_\mathrm{bin}$ smaller than about 3 AU (\cite{2007ApJ...658.1289T,2008ApJ...674.1086T,2008arXiv0808.1765Z}).

As first pointed out by \citet{2004A&A...417..353E}, 
{see also these proceedings},
there is statistical evidence for two interesting features in the mass-period and eccentricity
period distribution of planets residing in binary systems: \langeditorchanges{planets} with periods smaller
than about 40 days tend to have larger masses than their counterparts in single star systems,
while at the same time their eccentricities are smaller.
This trend has been supported by \langeditorchanges{the} more recent findings of \citet{2007A&A...462..345D} who tried
to correlate this with the tightness of the \langeditorchanges{binary,} but the statistics 
\langeditorchanges{are still based on small sample sizes} and 
more data are required.

As the influence of the secondaries on the planet formation process will obviously 
be smaller for larger distances, we shall focus in this contribution on the more challenging
tighter binaries and have used the physical \langeditorchanges{parameters} of the $\gamma$ Cep system for our
models. Interestingly, $\gamma$ Cep was one of the very first \langeditorchanges{stars} which has
been suggested to contain an extrasolar planet (of 1.7 $M_\mathrm{Jup}$):
{\it ``This star has the firmest evidence of a very low mass companion''}
\citep{1988ApJ...331..902C}. A statement
unfortunately retracted later by the same team \citep{1992ApJ...396L..91W}, only
to be rediscovered by \citet{2003ApJ...599.1383H}.
Today, this system is one of the tightest binary system known to contain a Jupiter-sized 
protoplanet.
For this reason, it has attracted much attention
in past years. Several studies looked at the stability and/or the
possibility of (additional) habitable planets in the system
\citep{2004RMxAC..21..222D,2004MSAIS...5..127T,2006ApJ...644..543H,2006MNRAS.368.1599V}.
In our studies we have taken the data for $\gamma$~Cep from \citet{2003ApJ...599.1383H}.
The more recent data by \citet{2007A&A...462..777N} \langeditorchanges{only slightly} change the dynamical status.
\section{Constraints on the planet formation process in binary \langeditorchanges{star systems}}
\label{kley-sec:constraints}
In a binary star system, the tidal torques of the companion generate strong spiral arms in the
circumstellar disk of the primary and angular momentum will be transferred
to the binary orbit which in turn leads to a truncation and restructuring of the disk.
The truncation radius $r_t$ of the disk depends on the binary separation $a_\mathrm{bin}$, its
eccentricity $e_\mathrm{bin}$, the mass ratio $q = M_2/M_1$ (where $M_1$, $M_2$ denote
\langeditorchanges{the} primary and secondary mass, respectively), and the viscosity $\nu$ of the disk.
For typical values of $q \approx 0.5$ and $e_\mathrm{bin}=0.3$ the disk will be truncated
\langeditorchanges{at} a radius of $r_t \approx 1/3 a_\mathrm{bin}$ for disk Reynolds numbers of $10^5$
\citep{1994ApJ...421..651A,1999MNRAS.304..425A}.
For a given mass ratio $q$ and semi-major axis $a_\mathrm{bin}$ an increase in $e_\mathrm{bin}$ will
reduce the size of the disk while a large $\nu$ will increase the \langeditorchanges{disk's radius}.

Whether these changes in the disk structure have an influence on the likelihood
of planet formation in such disks has been a matter of debate.
However, the dynamical action of the secondary induces several consequences which
appear to be adverse to planet formation: {\it i}) it changes the stability properties of orbits
around the primary, {\it ii}) it reduces the lifetime of the disk, and
{\it iii}) it increases the temperature in the disk.

Using numerical hydrodynamical studies, \citet{2000ApJ...537L..65N} argued
that both main scenarios of
planet formation, i.e. {\it core accretion} and {\it gravitational instability}, are
strongly handicapped, because an eccentric companion may induce a periodic heating
of the disk up to temperatures above the sublimation point of solids.
Since the condensation
of particles as well as the occurrence of gravitational instability require
lower temperatures, planet formation will be made more difficult in both scenarios.
Numerical studies of the early planetesimal formation phase
in rather close binaries with separations of only 20\langeditorchanges{--}30 AU show that
it is indeed possible to form planetary embryos in such systems
\citep{2004RMxAC..22...99L,2005MSAIS...6..172T,2007ApJ...660..807Q}.
Clearly, the possibility of embryo formation will depend strongly on the binary orbital elements, 
i.e. $a_\mathrm{bin}$ and $e_\mathrm{bin}$ and its mass ratio $q$.

Already in ordinary planet formation around single \langeditorchanges{stars,} the lifetime of the
disk represents a limiting factor in the formation of planets from \langeditorchanges{the} disk.
It has been suspected that the dynamical action of a companion will limit
the lifetime of disks substantially and place even tighter constraints on
the possibility of planet formation.
However, a recent analysis of the observational data of disks in binary stars
finds no or very little change in the lifetimes of the \langeditorchanges{disks,}
at least for separations larger than about 20 AU \citep{2007prpl.conf..395M}.

In summary, in a binary star system the formation of planets is altered
and most likely is handicapped due to the dynamical action of the companion
and the subsequent change in the internal structure of the protoplanetary disks.
\section{Disk evolution in binary stars}
\label{kley-sec:disks}
In \langeditorchanges{the} first study on the evolution of embedded Jupiter type protoplanets in disks in 
binary stars it \langeditorchanges{was} found that \langeditorchanges{migration} and mass growth \langeditorchanges{occur} faster
in tighter binaries \citet{2001IAUS..200..511K}. Even though this finding is in rough 
agreement with the \langeditorchanges{aforementioned} statistical evidence from the mass-period and eccentricity-period
\langeditorchanges{distributions}, the simulations are unrealistic in the sense that they start from the
artificial condition of unperturbed initial disks.
However, before inserting the planet into the disk it is necessary to first relax the disk
in the binary to its equilibrium configuration in the presence of the secondary.
This makes sure that \langeditorchanges{the} 
planetary evolution is not spoiled by long term transients due to the dynamical influence
of the secondary star. In this section we present the result of this equilibration process
of disks in binaries without embedded \langeditorchanges{planets}, with more details laid out in \citet{2008A&A...486..617K}.

We have chosen a binary with parameters very similar to
the $\gamma$ Cephei system. Specifically, we use $M_1 = 1.59 M_\odot$, $M_2 = 0.38 M_\odot$,
$a_\mathrm{bin} = 18.5$~AU and $e_\mathrm{bin} = 0.36$, which \langeditorchanges{translate} into a binary period
of $P = 56.7$~\langeditorchanges{yr}. The primary star is surrounded by a flat circumstellar disk, where
the binary orbit and the disk all lie in one plane, i.e. they are coplanar. 
The typical dynamical timescale in the disk, the orbital period at a few AU, is substantially
shorter than the binary period, but in a numerical simulation the system's evolution
can only be followed on these short dynamical time scales. 
To simplify the simulations we assume that the disk is vertically thin and perform only
2D hydrodynamical simulations of an embedded planet in a circumstellar disk which
is perturbed by the secondary. The disk is assumed to be non-selfgravitating.
We assume that the effects of the intrinsic turbulence of the disk
can be described approximately through the viscous Navier-Stokes equations,
which are solved by a finite volume method (code {\tt RH2D}) which is second order in
space and time \citep{1999MNRAS.303..696K}.
Finally, we assume that the disk is locally isothermal where the ratio of the vertical thickness
$H$ to the distance $r$ from the primary is constant, \langeditorchanges{with} $H/r =0.05$.
For the viscosity an $\alpha$ type parameterization is used with $\alpha = 0.02$.

\begin{figure}[ht]
\begin{center}
\includegraphics[width=.72\textwidth]{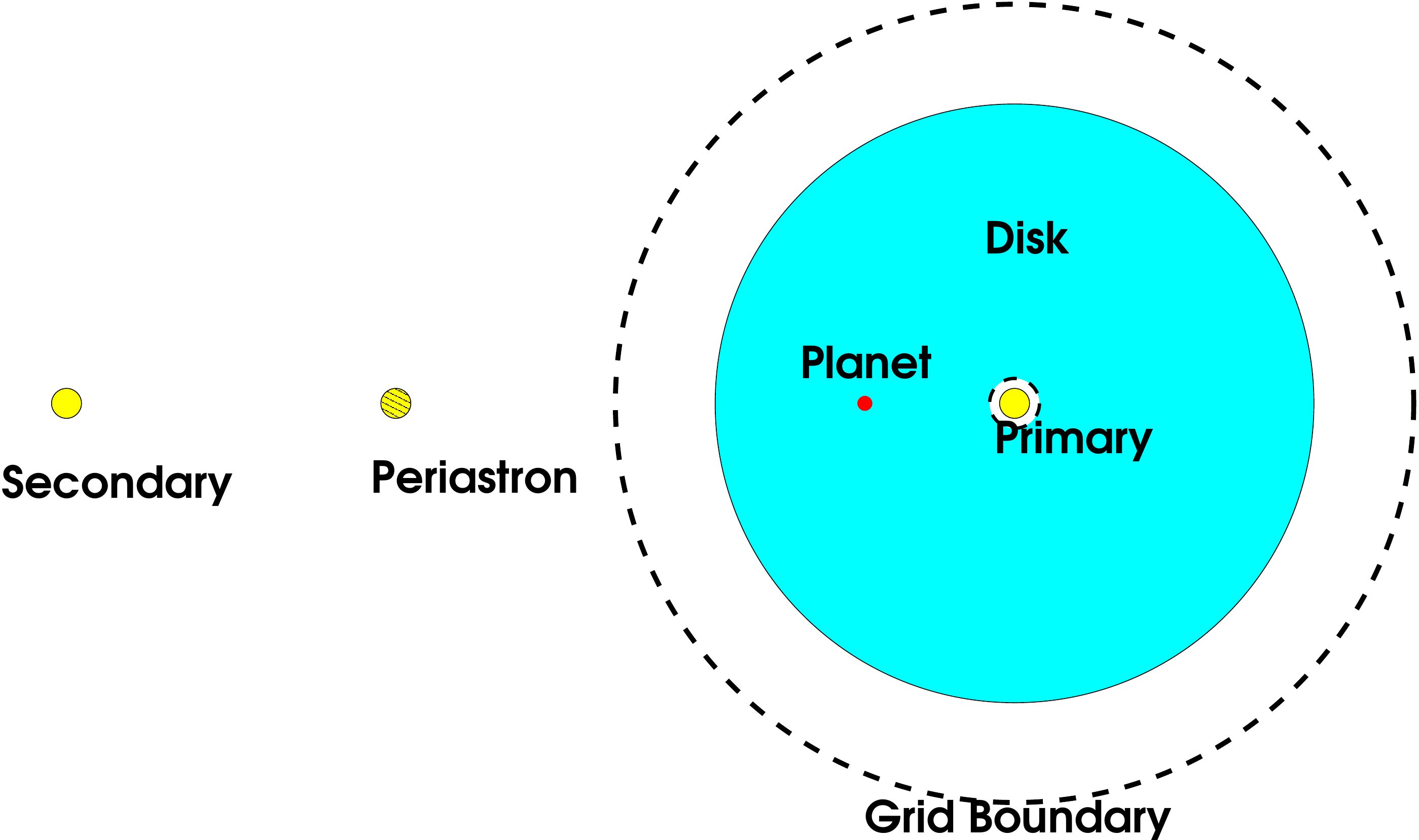}
 \caption{The numerical setup for simulations of disks in a binary \langeditorchanges{star system}.
 Here the binary \langeditorchanges{parameters} are $M_1 = 1.59 M_\odot$, $M_2 = 0.38 M_\odot$,
 $a_\mathrm{bin} = 18.5$~AU and $e_\mathrm{bin} = 0.36$. \langeditorchanges{For} the \langeditorchanges{disk,} the radial grid \langeditorchanges{extends}
 from $0.5$ to $8.0$ AU.
The left position of the secondary refers to the semi-major axis distance, the
right to the periastron.
}
   \label{kley-fig:setup}
\end{center}
\end{figure}
\begin{figure}[ht]
\begin{center}
\includegraphics[width=.48\textwidth]{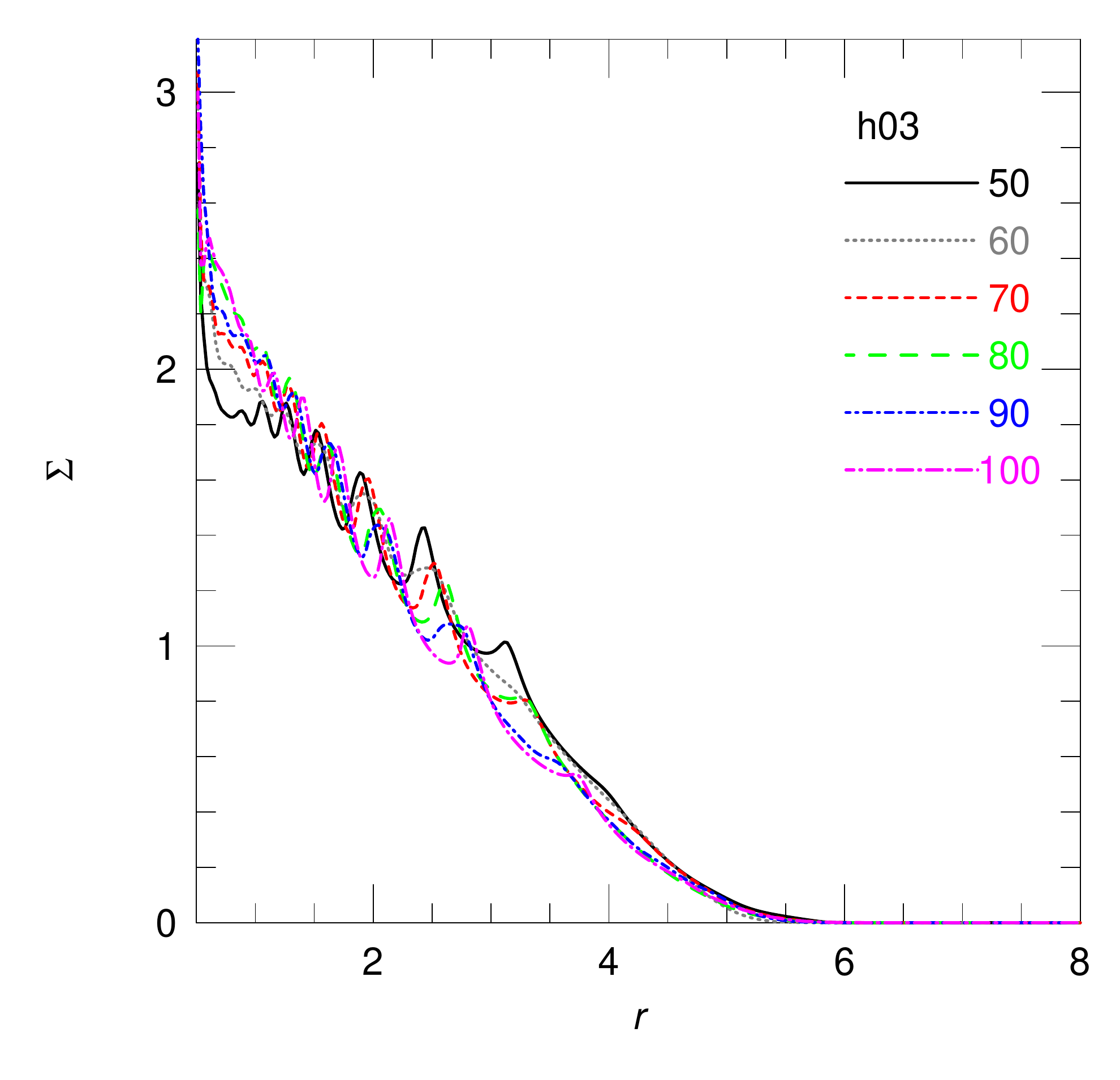}
\includegraphics[width=.48\textwidth]{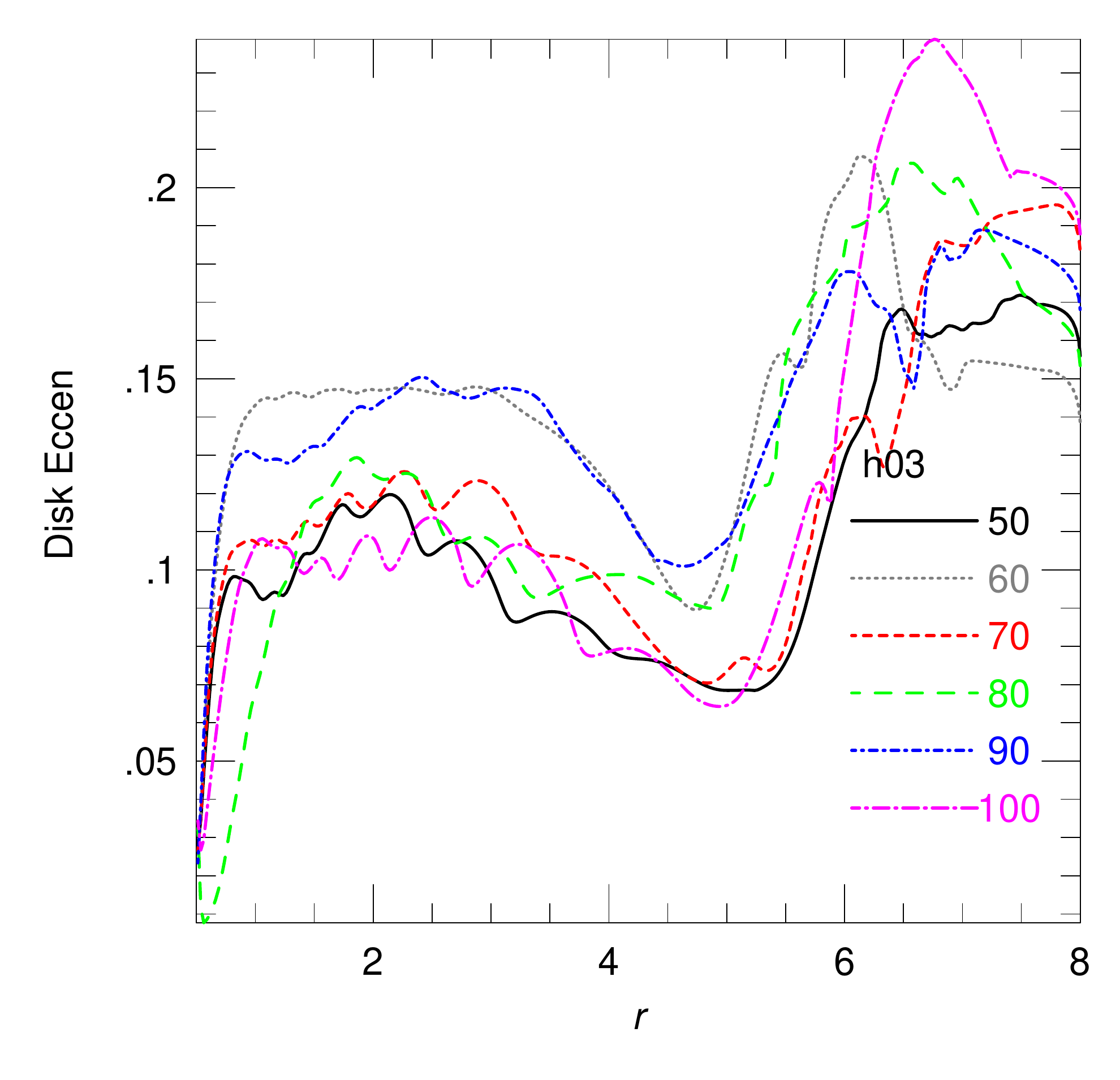}
 \caption{Azimuthally averaged disk structure at different evolutionary times given in binary orbits.
  On the left the azimuthally averaged surface density $\Sigma(r)$ is displayed and on the right
  the mean disk eccentricity at each radius, $e_\mathrm{disk}(r)$ \langeditorchanges{is shown}.}
   \label{kley-fig:diskstruct1}
\end{center}
\end{figure}
\begin{figure}[ht]
\begin{center}
\includegraphics[width=.48\textwidth]{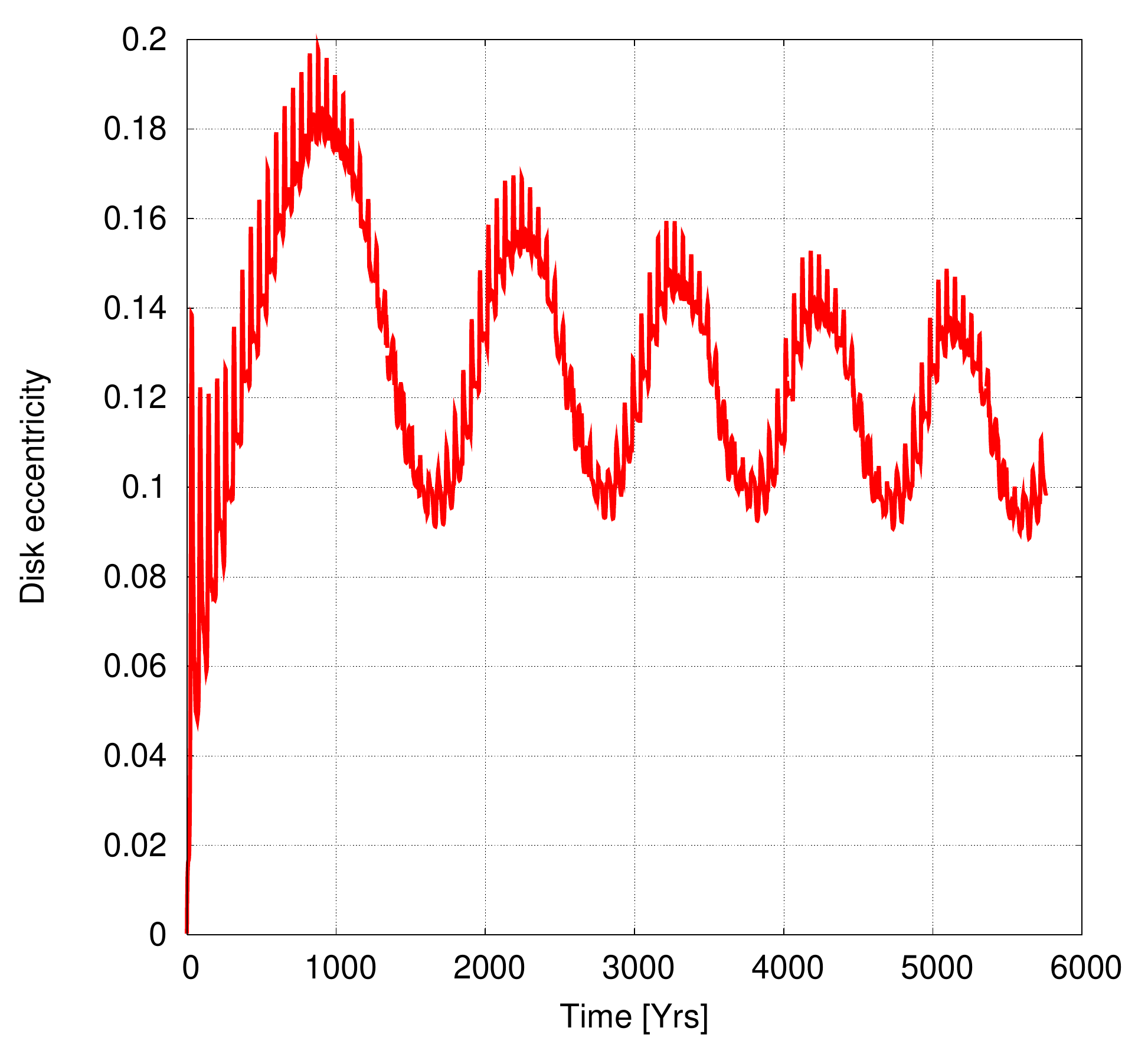}
\includegraphics[width=.48\textwidth]{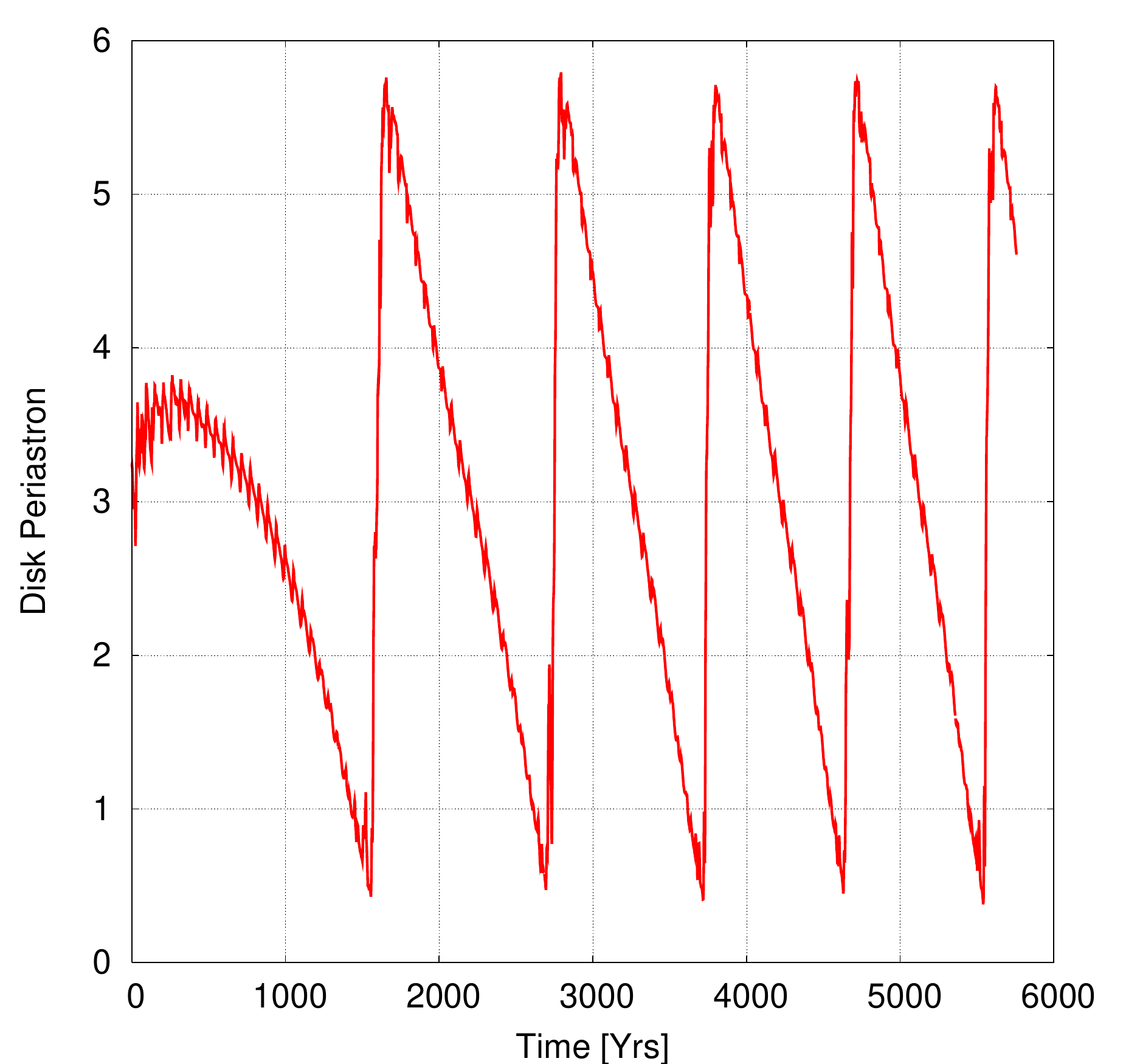}
 \caption{Global (mass averaged) disk eccentricity (left) and periastron (right) evolving in time.}
   \label{kley-fig:diskstruct2}
\end{center}
\end{figure}
In the runs presented \langeditorchanges{here} the computational domain covers a radial range from
$0.5$ to 8 AU, and 0 to $2 \pi$ in azimuth. This is covered with an
equidistant $300\times300$ grid. The numerical setup is displayed in Fig.~\ref{kley-fig:setup}.
To allow for parameter studies we have found it highly useful to increase
the performance of the code and have implemented the {\tt FARGO}-algorithm to
our code {\tt RH2D} which is \langeditorchanges{especially} designed to model differentially
rotating flows 
For our chosen radial range and grid resolution we find a speed-up
factor of about 7.5 over the standard case. Then, applying a Courant number of \langeditorchanges{0.75,}
about 160,000 timesteps are \langeditorchanges{still} necessary for only 10 binary orbits using our
setup, and we require hundreds of orbits.
\citep{2000A&AS..141..165M}.

In Fig.~\ref{kley-fig:diskstruct1} we display the end result of such an initial settling of
the disk in $\gamma$ Cep with no embedded planet.
The disk is truncated very
\langeditorchanges{early} in the simulations (in \langeditorchanges{fact,} after one or two binary orbits) and then re-adjusts as a whole
on longer, viscous timescales to reach equilibrium at around 60\langeditorchanges{--}70 binary orbits.
The disk is still perturbed periodically at each orbit due to the eccentric
orbit of the secondary. At around each periastron strong spiral arms appear in the disk
which are then damped until apoastron. However,
the azimuthally averaged density structure at $t=80$ and $t=100$ \langeditorchanges{no longer changes.} 
During the process of equilibration the average eccentricity of the disk, $e_\mathrm{disk}$,
settles to a value of about \langeditorchanges{0.1--0.15} in the \langeditorchanges{most massive} part of the disk. 
\langeditorchanges{The eccentricity remains high only}
in the \langeditorchanges{outer,} 
low\langeditorchanges{-}density parts of the \langeditorchanges{disk, where this is caused by} 
the secondary. 

In Fig.~\ref{kley-fig:diskstruct2} the time evolution of the global disk eccentricity and periastron
are displayed. In the presence of the companion the disk attains a non-zero eccentricity which
oscillates about 0.12 and the disk as a whole experiences
a coherent slow retrograde precession with a period of about 700 \langeditorchanges{years}
or 14 binary orbits.
The generation of a finite, non-zero disk eccentricity and precession is not restricted
\langeditorchanges{to the existence of an} eccentric binary orbit but \langeditorchanges{also} occurs in circular 
\langeditorchanges{systems,} driven
by an eccentric disk instability \citep{1991ApJ...381..259L}.
The conditions for the eccentricity growth have been analyzed for a wider parameter range
more recently by \citet{2008A&A...487..671K}. The direction and rate of the disk precession
\langeditorchanges{are} determined by the disk temperature, i.e. by the (relative) scale height $H/r$.
With respect to the $\gamma$ Cephei system this feature
has been described by  \citet{2008MNRAS.386..973P,2008A&A...486..617K}.

\section{Evolution of protoplanets in disks}
\label{kley-sec:indisks}

\begin{figure}[ht]
\centerline{\includegraphics[width=.99\textwidth]{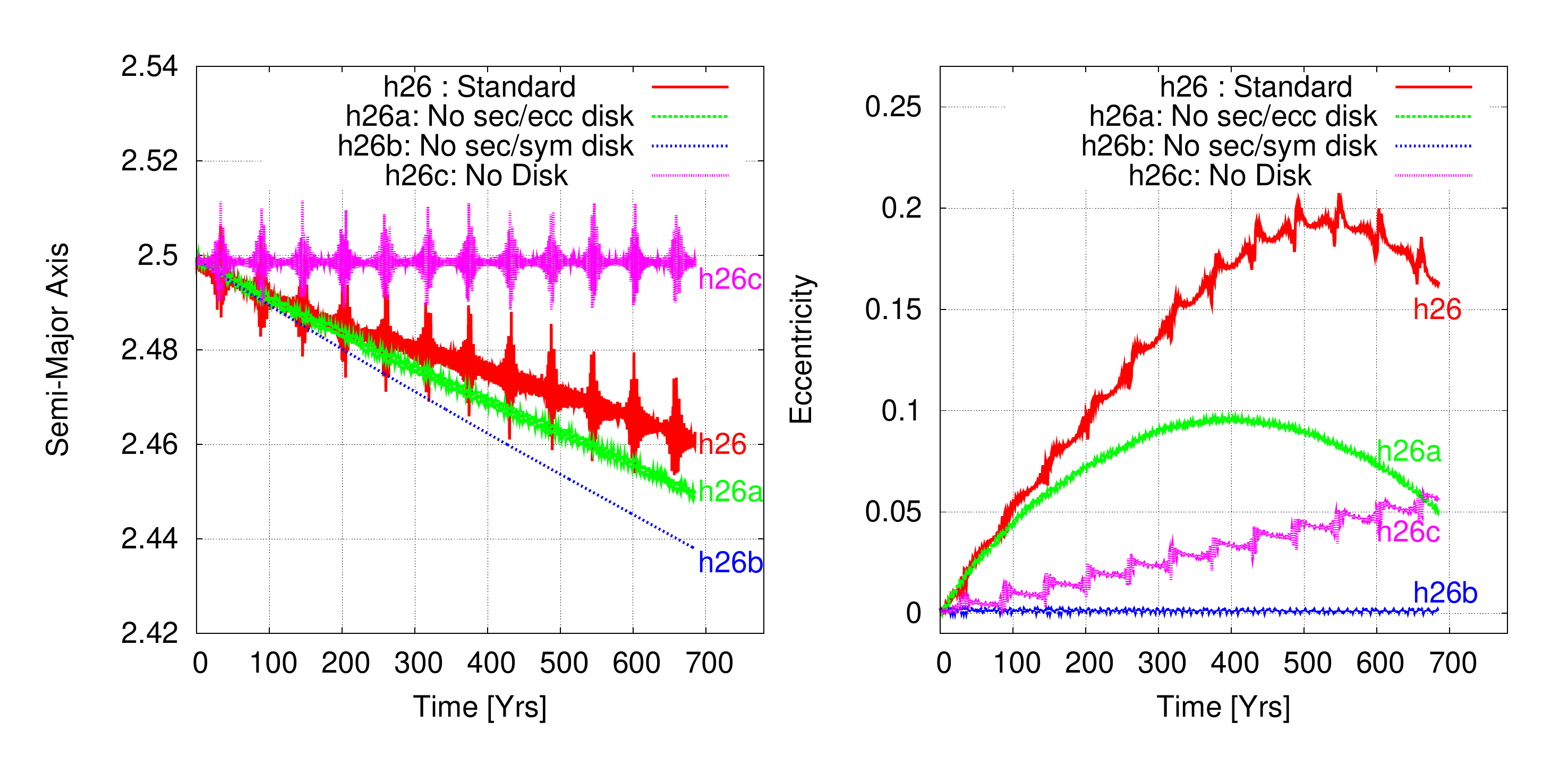}}
 \caption{Evolution of the semi-major axis and eccentricity for fiducial models where either the disk
 or the binary have been switched off individually to test their influence separately.
 The models that include the eccentric disk (here {\tt h26}: standard; {\tt h26a}: no secondary
eccentric disk) display the fastest growth in the planetary eccentricity.
  }
   \label{kley-fig:h26b-effect}
\end{figure}

The final two-dimensional density structure of the above equilibration process
(here at time $t=100$ binary orbits) is then used as the
initial condition for the embedded protoplanet simulations. The total mass of the disk
is rescaled to 3$M_\mathrm{Jup}$ and the planet of 36$M_\mathrm{Earth}$ is placed on a circular orbit
at a given semi-major axis (distance) from the primary star, ranging from $2.0$AU to $3.5$AU.

After inserting the protoplanet on a circular orbit at 2.5 AU we generally  expect that,
in addition to the typical planet-disk interaction,
its orbital elements will change due to the gravitational influence of the
binary and the distorted disk.
To differentiate the different contributions we have decided to check
the origin of the dynamical behavior for a non-accreting planet,
through a variation of physical conditions.
The standard model resembles the true physical situation where the
planet feels the full influence of the binary and the disk which is perturbed
by the binary. In the other setups we switch the various contributions on and
off.
The results, displayed in Fig.~\ref{kley-fig:h26b-effect},
show that the main contributor to the initial growth of planetary
eccentricity $e_\mathrm{p}$ is the eccentric disk and clearly not the eccentric binary,
for more details see \citet{2008A&A...486..617K}.

\begin{figure}[ht]
\centerline{\includegraphics[width=.99\textwidth]{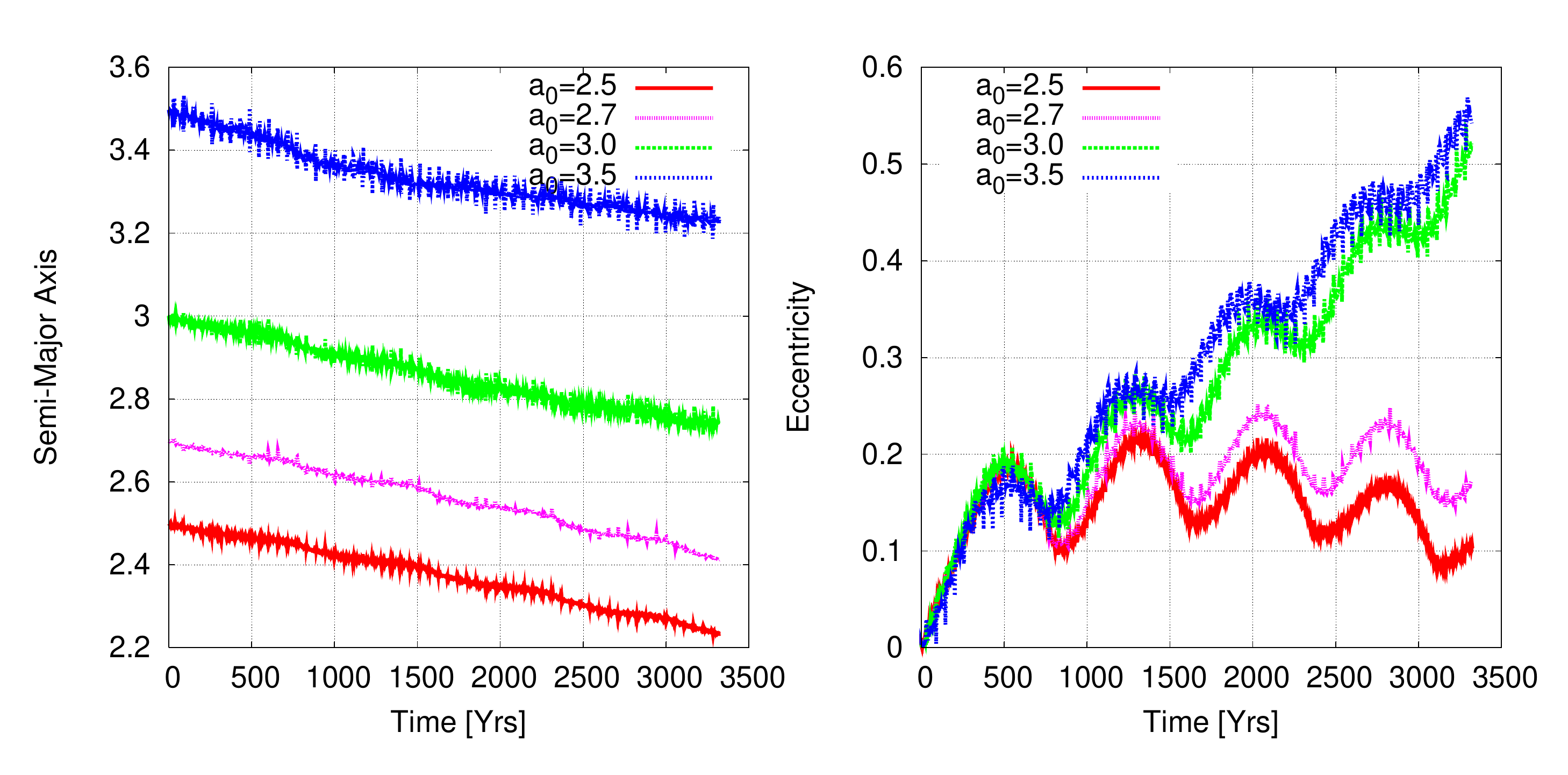}}
 \caption{The evolution of the semi-major axis and eccentricity for planets released at different
 distances from the binary, i.e. 2.5, 3.0 and 3.5 AU.
}
   \label{kley-fig:h28b-distance}
\end{figure}
\subsection{Evolving planets without mass accretion}
Planetary cores form in the outer cooler regions of protoplanetary disks beyond
the so called ice-line. However, in a binary star system the outer disk is affected most
by the secondary, and to find possible restrictions on the planet forming regions
in the disk it is important to analyze the evolution of cores near
the outer parts of the disk.
To study the effect of initial position we start our embryos at different locations
in the disk between 2.5 and 3.5 \langeditorchanges{AU,} always on a circular orbit, and \langeditorchanges{again} choose 
non-accreting cores. Because the initial characteristic growth time of the cores may be long,
even in comparison to the orbital period of the binary,
\langeditorchanges{this} set of runs \langeditorchanges{constitutes} a test suite to estimate the orbital evolution
of small protoplanets in the disk. 
The results for the semi-major axis and eccentricity evolution of the $36 M_\mathrm{Earth}$ planet
are displayed in Fig.~\ref{kley-fig:h28b-distance}, where the only difference in the four
cases is the release distance of the planet.
From all four locations the planet migrates inwards at approximately the same rate
with the tendency for a slow down for the two outer cases.
However, the different initial starting radii lead to a very different eccentricity
evolution.  Only the two innermost cases (starting at 2.5 and 2.7 AU) show \langeditorchanges{weak} 
eccentricity evolution,
the two outer cases display a very strong increase in their eccentricity beyond
$e_\mathrm{p}=0.5$ after about 55 binary orbits.
\langeditorchanges{Clearly,} the strongly disturbed disk in the outer regions at around 4~AU significantly
perturbs the orbits of the protoplanet and \langeditorchanges{initially} induces such high eccentricities
that the resulting elongated orbits \langeditorchanges{successively} become more and more influenced by the action of the binary.  
This increases the eccentricities to such high values that the orbits will 
\langeditorchanges{eventually become} 
unstable. The region of stability in this orbital domain has been
analyzed through simple \langeditorchanges{$N$}-body simulations \citep{2004RMxAC..21..222D,2004MSAIS...5..127T},
which match the results displayed here \langeditorchanges{very well}.

As the planets move on non-circular orbits in an eccentric disk and binary, a
temporal change of the apsidal line may be expected. 
However, the planets do not show a periastron precession but have
a stationary orientation \langeditorchanges{instead,} with some small oscillations of the periastron angle
about the mean with the same period as the oscillations in the eccentricity.
The innermost planet has a phase shift of \langeditorchanges{approximately} $180\deg$ with respect to
the binary and is nearly in an anti-symmetric \langeditorchanges{state,} while the other planets are
lagging behind this configuration \citep{2008A&A...486..617K}.
\begin{figure}[ht]
\centerline{\includegraphics[width=.66\textwidth]{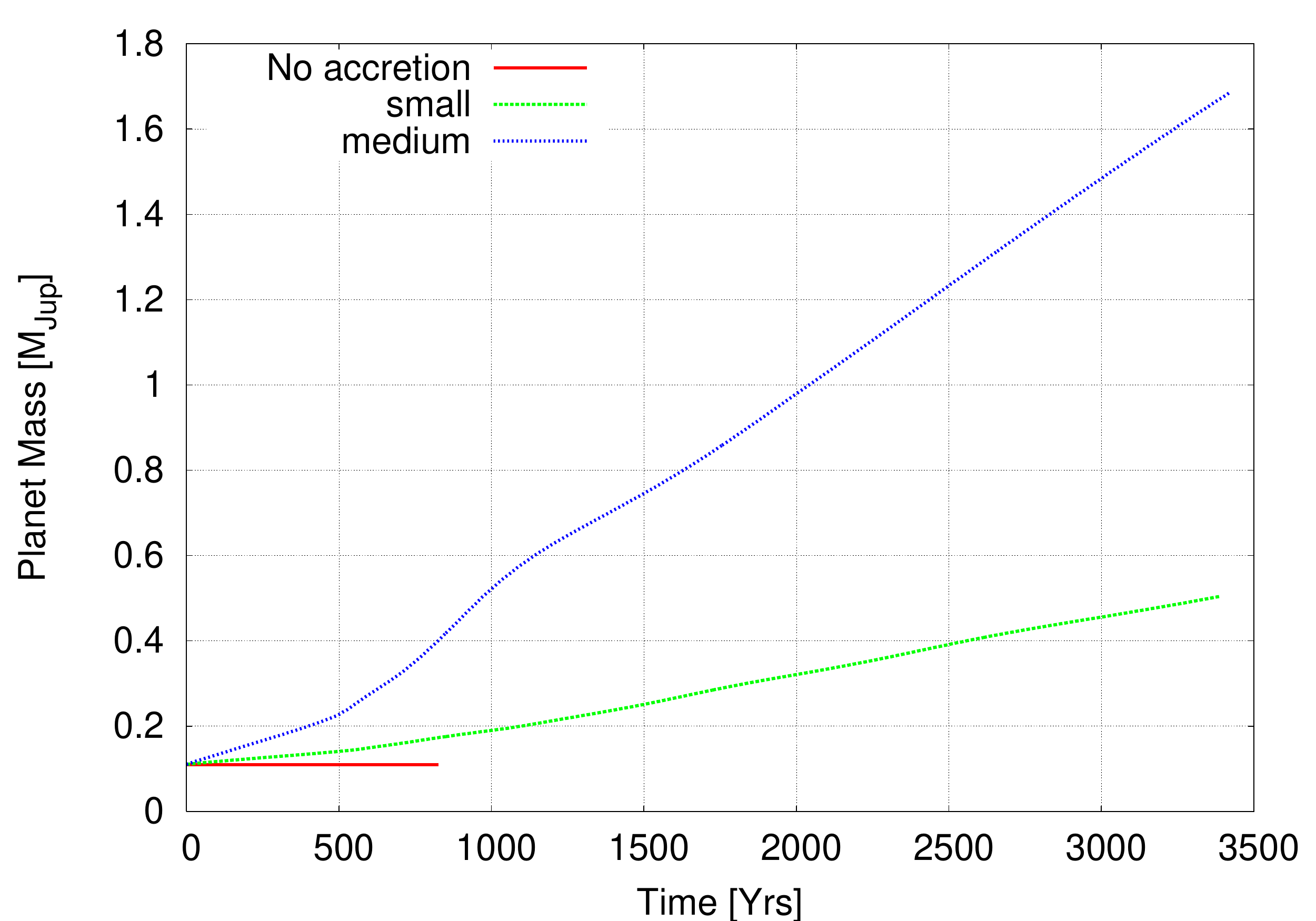}}
 \caption{Mass growth of a protoplanet 
  released at an initial distance of 2.3~AU after having evolved with constant mass
  (36 $M_\mathrm{Earth}$) from 2.5 AU to this location, see Fig.~\ref{kley-fig:h28b-distance}.
}
   \label{kley-fig:mp-h48}
\end{figure}
\begin{figure}[ht]
\centerline{\includegraphics[width=.99\textwidth]{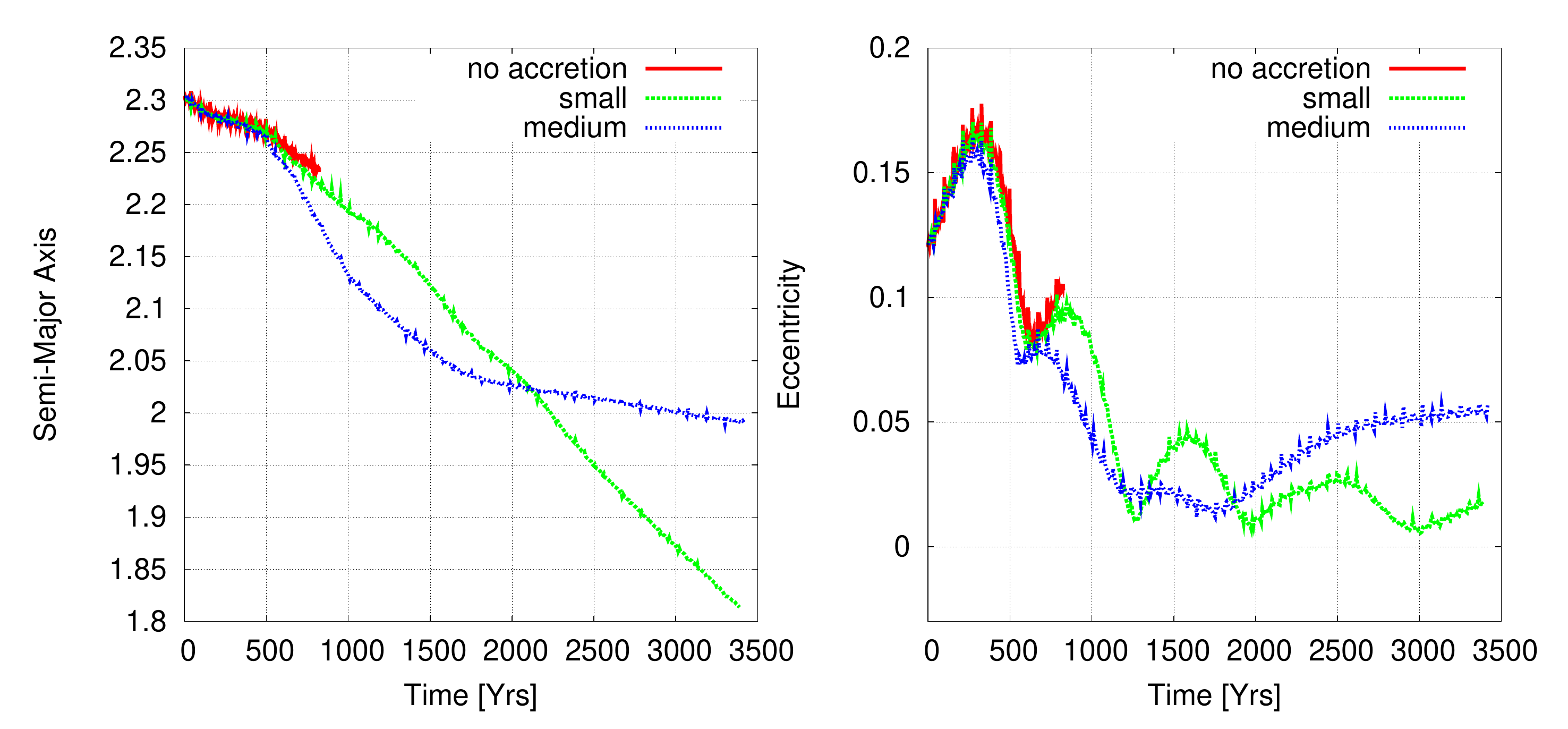}}
 \caption{The evolution of the semi-major axis and eccentricity for planets 
  released at an initial distance of 2.3~AU as in Fig.~\ref{kley-fig:mp-h48}.
}
   \label{kley-fig:aep-grow} 
\end{figure}
\subsection{Evolution with mass accretion}
To estimate the influence of protoplanetary accretion \langeditorchanges{on} the orbital evolution we have
\langeditorchanges{simulated} models where the mass of planets is allowed to grow due to accretion from
the ambient disk. This accretion process is modelled by taking out  
mass within a given radius $r_\mathrm{acc}$ from the Roche-lobe of the planet with different rates,
for detail see \citep{1999MNRAS.303..696K}.
For the medium accretion rate the mass of the planet reaches
about $1.6 M_\mathrm{Jup}$ after 3200 \langeditorchanges{yr} while the other models take \langeditorchanges{longer}.

The migration rate is initially similar for all accretion rates but then
accelerates as the mass of the planet increases (left panel of Fig.~\ref{kley-fig:aep-grow}),
and finally \langeditorchanges{slows} down because the mass reservoir of the disk becomes exhausted.
For the same reason (faster reduction of disk material) the final eccentricity of the
planet is smaller for higher accretion rates. Hence, the detailed evolution of the orbital
elements of the planet depends on the rate of mass accretion onto the planet.
The \langeditorchanges{efficiency} of the accretion process cannot be determined \langeditorchanges{straightforwardly,}
but is given for example by thermal processes in the vicinity of the growing planet.
In our simulations we did not find a single case of outward or highly reduced migration
\langeditorchanges{among the cases} of smaller planetary masses.
These assumed accretion rates are certainly much higher than realistic \langeditorchanges{ones,}
but they \langeditorchanges{provide an upper limit to how mass accumulation influences the}
orbital properties of growing planets.
The migration rate may also be affected by thermal processes in the disk.

\begin{figure}[ht]
\centerline{\includegraphics[width=.76\textwidth]{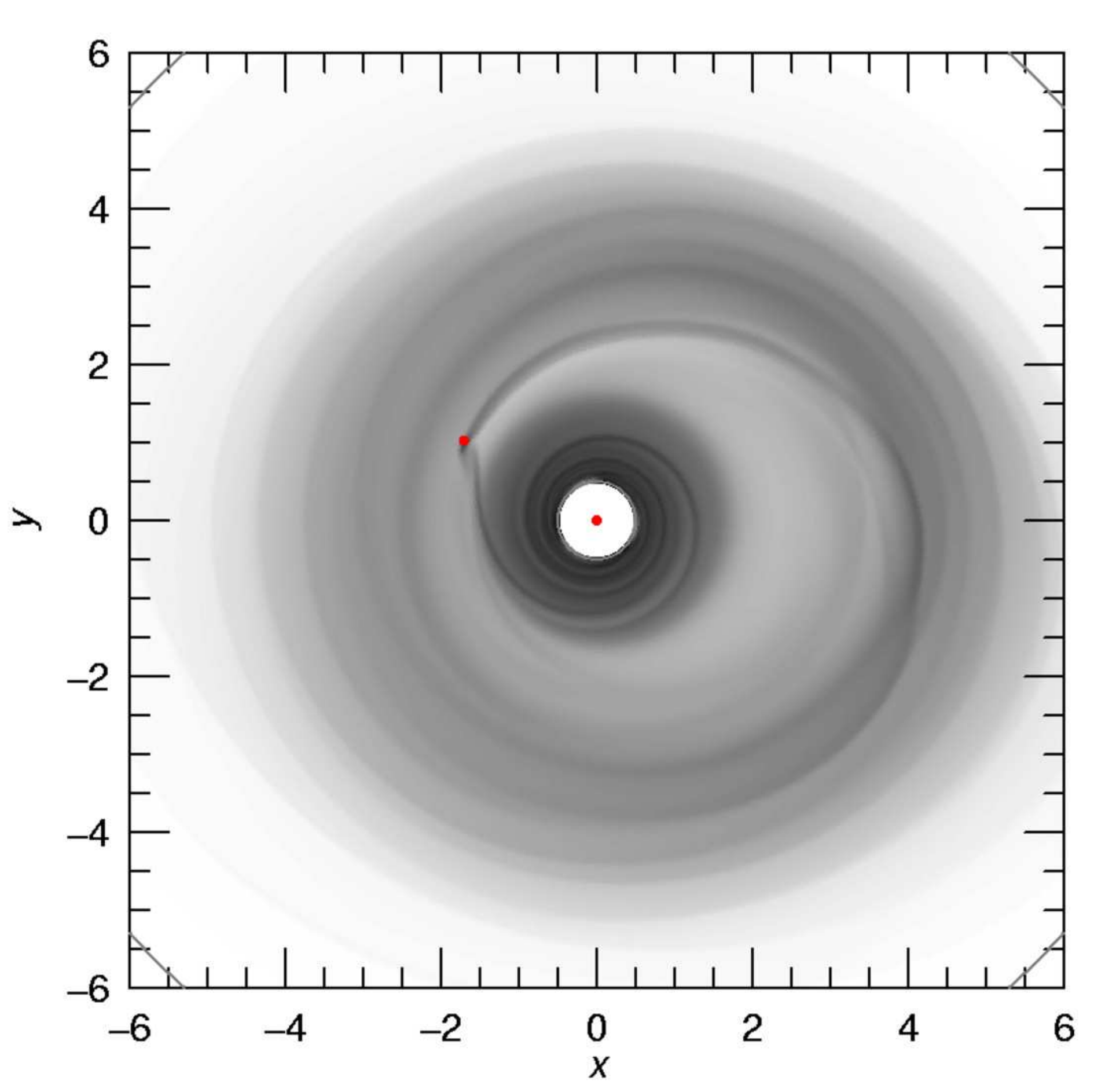}}
 \caption{
Grayscale plot of the two\langeditorchanges{-}dimensional density
  distribution of the medium accretion model at time
  $3125$~\langeditorchanges{yr}.  The
 shading is scaled $\propto \Sigma^{1/4}$ between $4.8 \times 10^{-4}$
 (white)   and $2400$ g/cm$^{2}$ (black).
  The location of the planet is marked by \langeditorchanges{a} small circle.
}
  \label{kley-fig:h48-sig2d} 
\end{figure}

A massive embedded planet will open a gap in standard circular disks, and it
is interesting to analyse this effect within the present context.
In Fig.~\ref{kley-fig:h48-sig2d} we display the two-dimensional density
distribution $\Sigma(r,\varphi)$ in the disk
at a time $3125$~\langeditorchanges{yr} for the medium accretion model. At this time the planet has reached
a mass of \langeditorchanges{approximately} 1.5 $M_\mathrm{Jup}$. From the plot it seems that the disk
inside the planetary orbit is apparently more circular than outside.
This is confirmed by the corresponding one-dimensional radial distribution of
the azimuthally averaged density and eccentricity of the disk at the same time.
The gap is somewhat weaker than in circular \langeditorchanges{disks,} primarily due to the periodic
disturbance of the secondary that tends to sweep material into the cleared region
around the planet.
Due to the shallower gap the planet is able to continue mass accretion
from its surroundings more easily \langeditorchanges{than} a planet on a circular
orbit in a single star system.
The inner disk clearly has a \langeditorchanges{lower} eccentricity than the outer \langeditorchanges{parts.}
The presence of the planet \langeditorchanges{represents,} in a \langeditorchanges{sense,} a barrier for the
(spiral) wave induced by the \langeditorchanges{binary,} which consequently cannot propagate
into the inner parts of the disk.
\section{Summary}
In this contribution we have concentrated on the planetary growth process in 
relatively tight binary stars with particular \langeditorchanges{attention given} to the
system $\gamma$~Cep.
To study the effect of the binary we have followed the
evolution of planetary embryos interacting with the ambient protoplanetary \langeditorchanges{disk,}
which is perturbed by the secondary star.

As suspected, the perturbations of the disk, in particular its
non-zero eccentricity and the periodic creation
of strong tidally induced spiral density arms, lead to non-negligible \langeditorchanges{effects} on
the planetary orbital elements. 
While embryos placed in the disk at different initial distances
from the primary star continue to migrate inwards at approximately the same rate, the
eccentricity evolution is markedly different for the \langeditorchanges{different} cases.
If the initial distance is beyond about $a \gsim 2.7$ AU the eccentricity of the embryo
continues to rise to very high \langeditorchanges{values,} and apparently 
\langeditorchanges{the orbit remains bound 
only due to the damping action of the
disk.} 
The main excitation mechanism of the initial rise of the eccentricity is
the perturbed disk and the spiral arms near the outer edge of the disk.
 
For a disk mass of 3$M_\mathrm{Jup}$ a $1.6 M_\mathrm{Jup}$ planet can easily be grown, and the final
semi-major axis and eccentricity are also in the observed range of the $\gamma$~Cep planet
for suitable accretion rates onto the planet.
One of the major problems in forming a planet in such a close binary system via the
core instability model is the problem of the formation of the planetary core in the
first place. Due to the large relative velocities induced in a planetesimal
\langeditorchanges{disk,} especially for objects of different \langeditorchanges{sizes,} the growth process is also problematic
in itself.

Hence, the formation of the \langeditorchanges{Jupiter-sized} planet observed in $\gamma$~Cep via the
standard scenarios remains difficult but may not be impossible. Future research will have
to concentrate on additional physical effects such as radiative transport, three-dimensional
effects and self-gravity of the disk.
\section{Acknowledgements}
This review is based on joint work with Richard Nelson from London (GB).
The work was sponsored in part by grant KL~650/6 of the German Research Foundation
(DFG).
\bibliographystyle{astron}

\end{document}